\documentclass[prl,twocolumn,10pt,aps,longbibliography]{revtex4-1}
\usepackage{verbatim}
\usepackage[english]{babel}
\usepackage{graphicx}
\usepackage{color}
\usepackage{amsmath}
\usepackage{soul}

\begin{document}
	
	\title{Observation of anomalous non-Ohmic transport in current-driven nanostructures}
	
	\author{Guanxiong Chen$^1$}
	\author{Ryan Freeman$^1$}
	\author{Andrei Zholud$^1$}
	\author{Sergei Urazhdin$^1$}
	
	\affiliation{$^1$Department of Physics, Emory University, Atlanta, GA, USA.}

	\begin{abstract}
Sufficiently large electric current applied to metallic nanostructures can bring them far out-of-equilibrium, resulting in non-Ohmic behaviors characterized by current-dependent resistance. We experimentally demonstrate a linear dependence of resistance on current in microscopic thin-film metallic wires at cryogenic temperatures, and show that our results are inconsistent with common non-Ohmic mechanisms such as Joule heating. As the temperature is increased, the linear dependence becomes smoothed out, resulting in the crossover to behaviors consistent with Joule heating. A plausible explanation for the observed behaviors is the strongly non-equilibrium distribution of phonons generated by the current. Analysis based on this interpretation suggests that the observed anomalous current-dependent resistance can provide information about phonon transport and electron-phonon interaction at nanoscale. The ability to control the properties of phonons generated by current can lead to new routes for the optimization of thermal properties of electronic nanodevices.

	\end{abstract}
	
\maketitle

Downscaling of modern electronic devices and circuits places ever increasing demands on their operation under increasingly non-equilibrium conditions. At small electric current density $J$ in metallic materials, the electric field $E$ is given by the Ohms law $E=\rho J$, where the material resistivity $\rho$ is assumed to be independent of $J$. At sufficiently large $J$, $\rho$ generally becomes bias-dependent, resulting in a nonlinear (non-Ohmic) dependence of $E$ on $J$.

Non-Ohmic behaviors are common in non-metallic structures.  For instance, the conductivity of tunnel junctions is generally bias-dependent, due to the phonon-mediated inelastic tunneling~\cite{PhysRevB.55.11738}.
The non-Ohmic behaviors observed in disordered graphene are associated with the bias-assisted hopping transport of charge carriers~\cite{kaiser2009electrical}. In semiconductors, the non-Ohmic behaviors at large bias are manifested by the phenomenon of current saturation, which is associated with the onset of spontaneous emission of optical phonons~\cite{1478102}. A strongly non-equilibrium distribution of electrons formed at large bias can also result in purely electronic non-Ohmic contributions to transport properties~\cite{battiato2017boltzmann}.

In metallic structures, the most common mechanism of non-Ohmic behaviors is associated with electron scattering on phonons generated by current, generally resulting in an increase of $\rho$ with increasing current. In macroscopic systems, current-generated phonons inevitably thermalize, and the effects of current can be well described as an increase of temperature $T^{*}(J)$ characterizing the distributions of both electrons and phonons. This temperature increase is defined as Joule heating~\cite{zangwill2013modern,neff1991introductory}. The thermal energy, generated at a rate $w=\rho J^2$ per unit volume, is dissipated by the diffusion of electrons and phonons away from the heated region, as described  by the Fourier's heat diffusion equation~\cite{ansermet2019principles}.

Since most materials exhibit a significant variation of resistivity with temperature, its current-dependence is commonly utilized for the characterization of Joule heating. This approach is also often extended to nanostructures~\cite{cheng2015temperature,PhysRevLett.93.166603,PhysRevLett.110.147601}. However, recent studies have shown that in nanoscale systems, the electron and the phonon distributions may not be adequately described by a single current-dependent temperature~\cite{pop2006heat,pop2010energy}, because electrons and/or phonons can escape from the system before they thermalize. We broadly define this regime as the breakdown of the Joule heating approximation. For instance, if electrons can quickly escape and are sufficiently weakly coupled to phonons, their effective temperature can be significantly lower than that of phonons, resulting in complex nonlocal energy transfer processes between the two subsystems~\cite{giri2015mechanisms,ono2018thermalization, PhysRevLett.55.422,PhysRevB.49.5942,PhysRevLett.76.3806,tripathy2008evidence}. In this case, the electrons and phonons can still be separately characterized by effective temperatures, which are generally not the same for the two subsystems.

Electrons and/or phonons may also form a nonequilibrium distribution within the respective subsystem, which cannot be characterized by an effective temperature~\cite{klett2018relaxation,kash1985subpicosecond,del2000nonequilibrium}.  Significant progress has been recently achieved in the understanding of non-equilibrium states of electrons and phonons at nanoscale~\cite{PhysRevB.97.205412,PhysRevB.76.085433, chase2016ultrafast,PhysRevB.98.045104, Huberman375}. However, a comprehensive microscopic understanding of current-induced thermal energy generation and transport has not yet emerged. 

Here, we experimentally demonstrate that at cryogenic temperatures, common nanostructures such as thin-film metallic nanowires exhibit an anomalous linear variation of resistance with current. We show that this dependence cannot be described by a current-dependent temperature, and is thus inconsistent with the Joule heating approximation. As the temperature is increased, the linear dependence is broadened, and behaviors consistent with Joule heating eventually emerge at sufficiently high temperatures. Nevertheless, signatures of anomalous current dependence persist at temperatures as high as $200$~K. We attribute the observed behaviors to substantially non-equilibrium phonon distribution facilitated by the fast phonon escape from the nanoscale system. Based on our interpretation, we show that the observed anomalous behaviors can provide insight into the nonequilibrium electron and phonon dynamics at nanoscale, lead to new approaches to the characterization of electron-phonon interaction, and facilitate the optimization of thermal management in electronic nanodevices.

	\begin{figure}
		\includegraphics[width=\columnwidth]{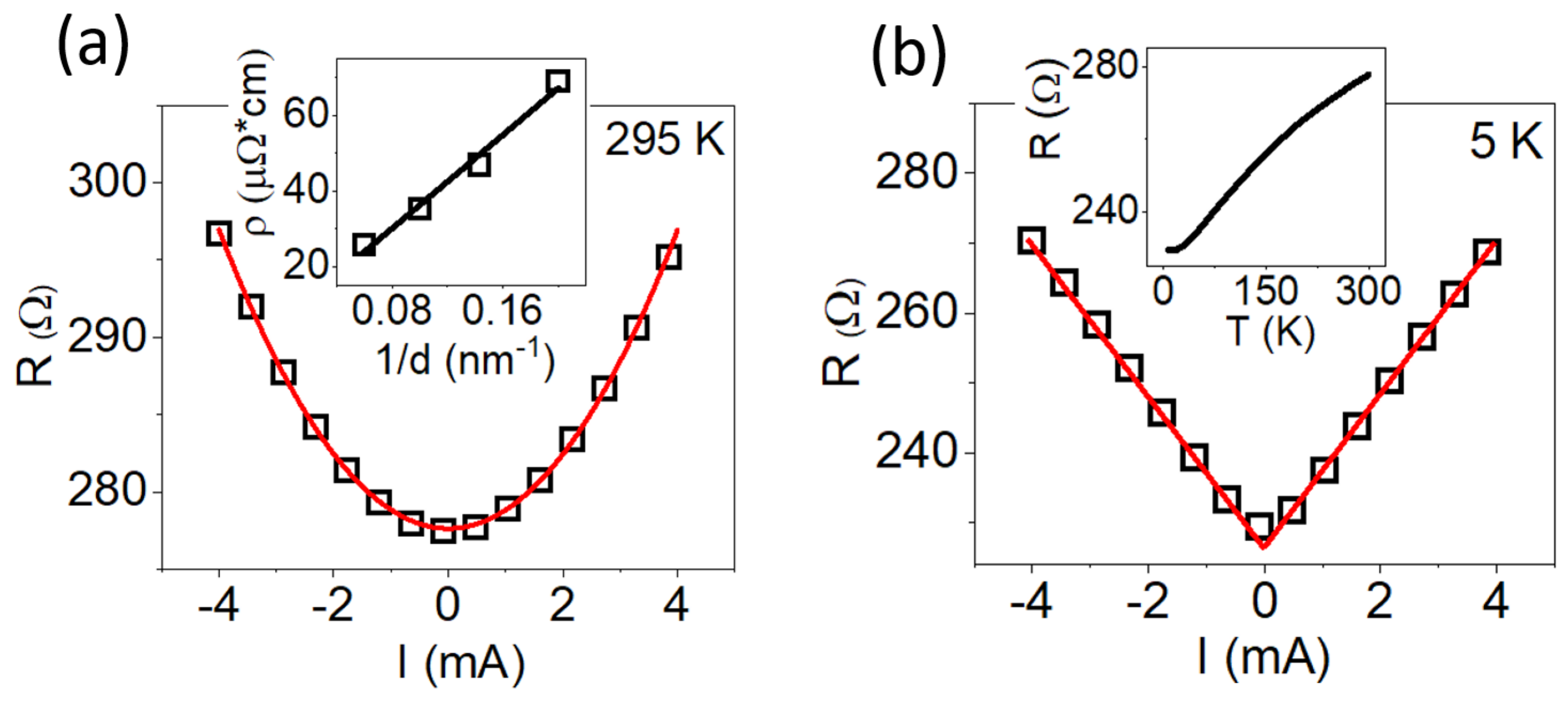}
		\caption{(Color online). (a) Resistance vs current for a $1$~$\mu$m-long, $500$~nm-wide, $5$~nm-thick Pt wire deposited on etched Si substrate, at room temperature $T=295$~K. Curve is the best fit to the data with a quadratic function. Inset: Resistivity of Pt deposited on Si substrates vs inverse Pt thickness, at $295$~K.	(b) Same as (a), at $T=5$~K. Curve  is a fit with the linear function $R(I)=R(0)+\alpha|I|$, where $\alpha$ is a fitting parameter. Inset: the dependence of the wire resistance on temperature at $I=0$.}\label{fig_1}
	\end{figure}
	
Below, we discuss mainly the results for thin-film Pt wires deposited on undoped Si substrates.  Similar wires are extensively utilized for spin current generation in spin-orbitronic devices that rely on the spin Hall effect exhibited by Pt~\cite{PhysRevLett.106.036601,jungwirth2012spin,PhysRevLett.101.036601,demidov2014nanoconstriction,yang2016spin}, as well as microscale heaters and thermometers~\cite{Kang2017}. We have also performed measurements for other materials and substrates: Au wires on Si, as briefly discussed below, Pt wires on SiO$_2$ and sapphire, as well as resistive metallic nanocontacts~\cite{supplementary}. All these measurements yielded consistent results, suggesting that the observed non-equilibrium phenomena are quite general to micro- and nano-scale current-driven structures characterized by efficient phonon relaxation. 

The studied wires were fabricated by a combination of e-beam lithography and high-vacuum sputtering. To ensure consistent thermal contact between the wires and the substrate, native surface oxide was removed from the Si surface by HF etching immediately prior to the Pt wire deposition. The deposited Pt was polycrystalline, due to the large lattice mismatch between Pt and Si. The wires were contacted by four Cu(150) electrodes for the four-probe resistance measurements, which allowed us to eliminate the contribution of the contact resistance between the electrodes and the wires. Here and below, numbers in parentheses are thicknesses in nanometers. Differential resistance $R=dV/dI$ was measured using the standard lock-in technique, with ac current $I_{ac}=10$~$\mu A$ rms superimposed with the dc current $I$ of up to $\pm 4$~mA.

The measured resistivity of the studied Pt films linearly depended on the inverse film thickness $d$ [inset in Fig.~\ref{fig_1}(a)], consistent with the expected contribution of surface scattering~\cite{Dutta2017}. For large thicknesses, the resistivity approaches the usual values reported for bulk sputtered Pt~\cite{Dutta2017}. We note that thin Pt films deposited on SiO$_2$ and sapphire exhibit significantly lower resistivity than similar films on Si~\cite{supplementary}. Since the roughness $\approx 0.25-0.3$~nm rms of the substrate and of the film surface, as measured by atomic force microscopy, was similar for all these films, we conclude that the large resistivity of thin Pt films on Si is caused by the diffuse electron scattering at the Pt/Si interface. This allowed us to explore a large range of transport parameters controlled by the film thickness. We emphasize that the reported nonequilibrium effects were also observed in thin films deposited on sapphire and SiO$_2$, where the interface contribution to resistivity was much smaller~\cite{supplementary}. Therefore, these effects are not associated with the specific scattering properties of the interface.

Figure~\ref{fig_1} shows representative resistance vs current curves for a $1$~$\mu$m-long and $500$~nm-wide Pt(5) wire. At the experimental temperature $T=295$~K, the resistance $R(I)$ follows a quadratic dependence on current $I$ [Fig.~\ref{fig_1}(a)]. This result is consistent with Joule heating. Indeed, electrical energy is dissipated in the wire at a rate $W=RI^2$. Since the rate of thermal energy dissipation from the wire is proportional to its temperature increase, one can expect that the temperature of the wire follows a quadratic dependence on current $T^{*}\approx T_{0}+CI^{2}$. The resistance of the Pt(5) wire exhibits an approximately linear dependence on $T$ close to $T=295$~K [inset in Fig.~\ref{fig_1}(b)]. Thus, $R(I)$ is expected to follow a quadratic dependence, in agreement with our data.  

At $T=5$~K, the dependence $R(I)$ is well described by the linear function $R(I)=R_0+\alpha|I|$, where $\alpha$ is a constant [Fig.~\ref{fig_1}(b)]. This dependence cannot be explained by Joule heating. In particular, numerical simulations confirm that for Joule heating, the dependence $T^{*}(I)$ of the Pt wire temperature on current should be quadratic even at cryogenic temperatures~\cite{supplementary}. The resistance of Pt is almost independent of temperature up to $20$~K, due to the freeze-out of large-momentum phonons, and starts to increase approximately linearly with $T$ at higher temperatures [inset in Fig.~\ref{fig_1}(b)]. Simulations show that as a consequence, Joule heating should not affect $R$ for currents up to about $1.8$~mA, and should lead to an approximately quadratic dependence $R(I)$ for larger currents~\cite{supplementary}. The observed dependence $R(I)$ is clearly qualitatively inconsistent with these expectations.

The anomalous current dependence of resistance cannot be attributed to tunneling or electron hopping effects discussed in the introduction, because the electron transport properties of the studied metallic thin film wires are determined by scattering rather than tunneling, as can be inferred from the decrease of resistance with decreasing temperature [inset in Fig.~\ref{fig_1}(b)]. The effects of contact resistance are eliminated in our measurements by the four-probe geometry. We also performed separate two-probe measurements that produced very similar results, except for the overall resistance increase due to the contribution of electrodes, confirming that the contact resistance is not relevant to the observed effects.

We can also eliminate the purely electronic  contribution to the dependence $R(I)$, because the linear term in this dependence does not change sign with the direction of current, and is negligible for inversion-symmetric materials such as Pt.~\cite{battiato2017boltzmann}. One can also expect that such effects should also sensitively depend on the band structure. However, our results for an Au wire discussed below are very similar to those for Pt, despite large differences in their band structure. Resistance may be also affected by the complex nonlocal transport phenomena, which have been reported for mesoscopic wires with mean free path comparable to the wire dimensions~\cite{Pothier1996}. However, we estimate that the mean free path in our Pt(5) film is only $2$~nm, due to the diffuse electron scattering at the Pt/Si interface, resulting in negligible nonlocal effects.

We hypothesize that the observed anomalous low-temperature dependence $R(I)$ is associated with electron scattering on non-equilibrium phonons generated by current, whose distribution and population are qualitatively different from that expected from the Joule heating picture. Indeed, analysis presented below indicates that the linear $R(I)$ dependence can be explained within the Drude-Sommerfeld approximation for electron transport, in conjunction with the simple kinetic analysis of phonon relaxation in the limit of negligible thermalization. Independent estimates of phonon relaxation rates confirm the validity of this approximation~\cite{supplementary}. Fast escape of the generated phonons from the system  is expected to be the only requirement for the observed non-equilibrium behaviors, and therefore such behaviors should be quite common in current-driven micro- and nano-structures characterized by efficient phonon relaxation.

 	\begin{figure}
	\includegraphics[width=\columnwidth]{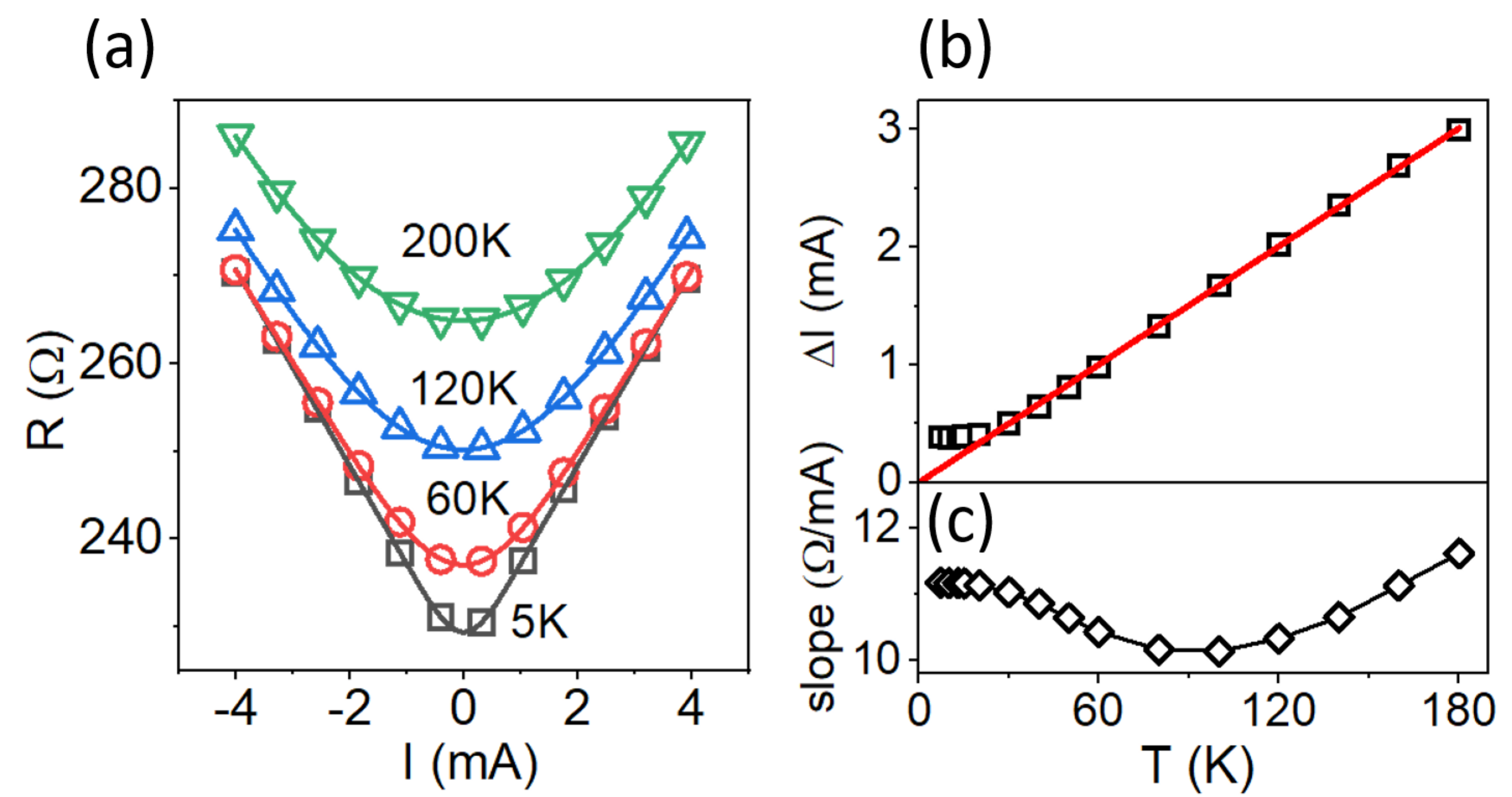}
	\caption{(Color online). 
		 (a) Symbols: $R$ vs $I$ for the same sample as in Fig.~\ref{fig_1}, at the labeled values of $T$. Curves: best fits with the linear function $R(I)=R_0+\alpha|I|$ convolved with the Gaussian $g(I)=\frac{1}{\sqrt{2\pi}\Delta I}e^{-I^2/2\Delta I^2}$. Some of the data points are omitted for clarity, but fitting was performed for the entire data set.  (b),(c) Parameters extracted from the data fitting: the Gaussian width $\Delta I$ (b) and the slope of the linear dependence (c). The line in (b) is the best linear fit of the data for $T>20$~K.
			}\label{fig_2}	\end{figure}

We now discuss the dependence on temperature, which provides insight into the mechanism of the emergence of behaviors consistent with Joule heating at high temperatures. As the temperature is increased from $5$~K, the linear dependence remains evident at large currents, but the zero-current singularity becomes increasingly smoothed-out [Fig.~\ref{fig_2}(a)]. This broadening is reminiscent of the thermal effects observed in electronic spectroscopy of tunnel junctions and point contacts, which can be accounted for by convolving the zero-temperature spectra with the normalized derivative $\frac{df_0}{d\epsilon}$ of the Fermi-Dirac distribution function $f_0(\epsilon)=\frac{1}{e^{\epsilon/k_BT}+1}$, where  $\epsilon $ is the energy relative to the chemical potential~\cite{Wiesendanger1994}.  This function can be well approximated by the Gaussian 
\begin{equation}\label{Gaussian}
g(\epsilon)=\frac{1}{\sqrt{2\pi}\sigma}e^{-\epsilon^2/2\sigma^2},
\end{equation}
with $\sigma\approx 1.6k_BT$. Indeed, a linear function $R(I)=R(0)+\alpha |I|$ convolved with the Gaussian 
\begin{equation}\label{Gaussian2}
g(I)=\frac{1}{\sqrt{2\pi}\Delta I}e^{-I^2/2\Delta I^2}.
\end{equation}
provides a good fitting for all our $R(I)$ data measured at different temperatures, as shown by curves in Fig.~\ref{fig_2}(a). At $T>200$~K, the width $\Delta I$ of the Gaussian becomes larger than the range of the dc current scan, resulting in a significant uncertainty of the fitting. Nevertheless, these data suggest that a non-equilibrium current-driven phonon distribution, not described by Joule heating, can be formed at sufficiently large bias even in the ambient temperature range. On the other hand, at small currents, thermal broadening results in a quadratic $R(I)$, consistent with the dependence expected for Joule heating.

We emphasize that in the studied microstructures, phonons are not expected to thermalize even at elevated temperatures, due to their efficient relaxation. Instead, behaviors consistent with Joule heating emerge because of the thermal broadening of the electron distribution, resulting in a nearly thermal distribution of the generated phonons.

In tunnel junctions or point contacts, the characteristic electron energy competing with the thermal broadening is defined by the bias across the junction~\cite{Wiesendanger1994}. We hypothesize that in the studied metallic wires, the characteristic electron energy defining the thermal broadening scale is the energy provided to electrons by electric field between the scattering events, which is to a good approximation proportional to the bias current. Indeed, the Gaussian width $\Delta I$ follows a linear dependence at $T>20$~K, extrapolating to $\Delta I=0$ at $T=0$ [right scale in Fig.~\ref{fig_2}(b)]. This result is consistent with our hypothesis that the observed broadening of $R(I)$ originates from the thermally induced spectral broadening of the electron distribution, whose width is proportional to temperature. In other words, the thermal broadening of $R(I)$ can be interpreted in terms of the competition between the thermal energy $k_BT$ of electrons, and the average energy acquired by electrons between scattering events due to the electric field in the wire. The broadening saturates at $T<20$~K, suggesting the existence of an additional non-thermal broadening effect. Indeed, at small bias and low temperatures, the energy (and thus the momentum) of phonons generated by current is small, resulting in a reduced contribution of the generated phonons to electron scattering, and thus resistance. We expect that detailed theoretical analysis of these broadening effects, which is beyond the scope of the present work, will provide quantitative information about the relevant energy scales, leading to new quantitative insights into electron transport in nanostructures.

The slope of the linear dependence decreases with increasing temperatures up to $90$~K, and then increases at higher temperatures [Fig.~\ref{fig_2}(c)]. These variations are likely associated with the temperature dependence of phonon relaxation rate, which is dominated by the dissipation into the substrate, as shown below. Qualitatively, there is some correlation between the observed variations of the slope and the temperature dependence of thermal conductivity of bulk Si, which increases with increasing temperature at low temperatures, exhibits a peak typically at temperatures between $30$~K and $100$~K, and then decreases at higher temperatures~\cite{glassbrenner1964thermal,asheghi1998temperature}. However, the variations of thermal conductivity are almost two orders of magnitude larger than those of the slope in Fig.~\ref{fig_2}(c), suggesting that phonon relaxation may not be adequately described by the diffusive heat transport. Indeed, the phonon mean free path in Si significantly exceeds the dimensions of the studied structures~\cite{gereth1964phonon}. Therefore, phonons are expected to quasi-ballistically escape into the Si substrate, with negligible contribution from the phonon-phonon scattering governing heat diffusion. Thus, measurements of temperature-dependent slope of $R(I)$ in nanostructures, such as those discussed here, may provide a new method for characterizing phonon transport and relaxation at nanoscale.
 
	\begin{figure}
	\includegraphics[width=\columnwidth]{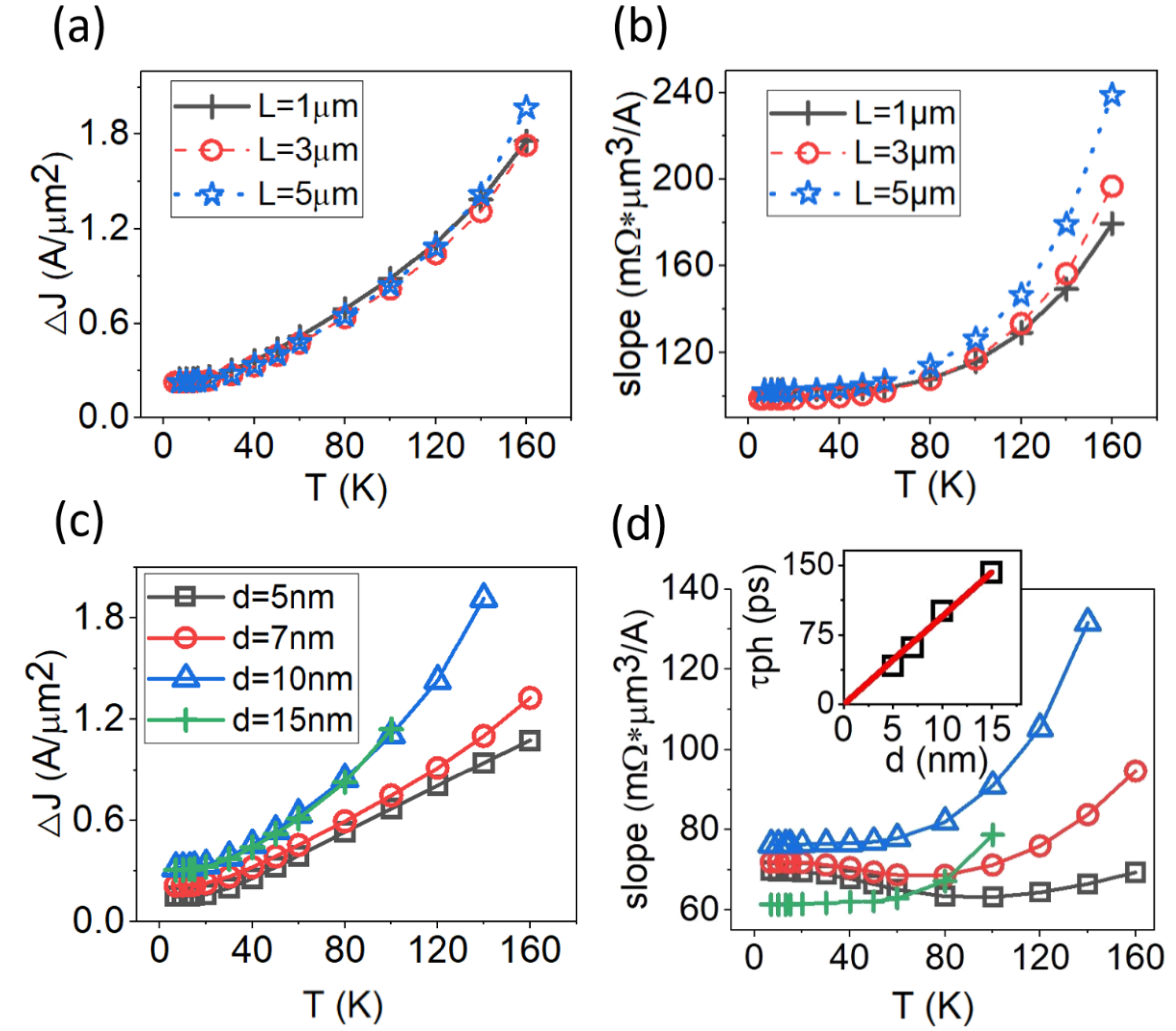}
	\caption{(Color online). 
		Effects of the Pt wire geometry. (a),(b) Temperature dependence of the Gaussian broadening width $\Delta J$ (a) and the linear slope (b) of $\rho(J)$, for $10$~nm-thick Pt wires with lengths $L=1$~$\mu$m, $3$~$\mu$m, and $5$~$\mu$m, as labeled. (c),(d) Same as (a), (b), for $1\mu$m-long Pt wires with thicknesses $d=5$~nm, $7$~nm, $10$~nm, and $15$~nm, as labeled in (c). Inset in (d): Dependence of the phonon relaxation rate $ \tau_{ph} $ on the Pt thickness, determined from the data at $T=5$~K using Eq.~(\ref{rholin}) (symbols), and linear fit with zero intercept (line). 
    }\label{fig_3}
	\end{figure}

Non-equilibrium phonon distribution in the studied Pt wires, manifested by the linear dependence $R(I)$, is associated with fast relaxation of phonons due to their efficient escape from the system, before they become thermalized. To elucidate the mechanisms of phonon relaxation, we analyze the effects of the wire geometry. Phonons are expected to dissipate mainly into the thick Cu leads and the substrate. If phonons dissipate predominantly into the leads, then the efficiently of their dissipation should decrease with increasing wire length $L$, due to the longer average distance they travel before escaping. On the other hand, the phonon dissipation efficiency, per unit Pt volume, should not be significantly affected by the wire thickness $d$, aside from the effects of thickness on the phonon mean free path. However, if phonons dissipate predominantly into the substrate, the length of the wire should not affect the phonon escape into the substrate. On the other hand, increasing the wire thickness $d$ should lead to an increase of the average phonon escape time, resulting in less efficient dissipation (per unit Pt volume).

Figure~\ref{fig_3} summarizes the results for different values of $L$ and $d$. To facilitate direct comparison of different geometries, we analyze the dependence of resistivity $\rho=RA/L$ on the current density $J=I/A$, where $A$ is the cross-section area of the wire. In all cases, we obtained accurate fits of the data by using a linear dependence convolved with the Gaussian. For the thickest (15 nm) Pt wire, the analysis is limited to $T\le 100$~K, because of the large thermal broadening for this thickness.

The thermal broadening $\Delta J$ is independent of the wire length [Fig.~\ref{fig_3}(a)]. The linear slope of $\rho(J)$ is almost independent of length at temperatures up to about $80$~K, and starts to increase with increasing length at higher temperatures. However, the dependence on length remains modest even at $T=160$~K. Thus, we conclude that at cryogenic temperatures, phonons relax in the studied Pt wires predominantly through the substrate. 

In contrast to the effects of the wire length, wire thickness significantly affects the characteristics of $\rho(J)$, Figs.~\ref{fig_3}(c),(d). The broadening increases by a factor of two when $d$ is increased from $5$~nm to $10$~nm, and saturates at larger $d$. For Pt(5), it increases linearly with temperature $T>20$~K. This result is consistent with our interpretation of the broadening in terms of the competition between the electron's thermal energy $kT$ and the energy $\propto J$ provided by the electric field between electron scattering events. The observed curving of the dependence $\Delta J(T)$ for larger Pt thickness, especially apparent in Fig.~\ref{fig_3}(c) for Pt(10) and Pt(15), can be attributed to the larger relative contribution of electron-phonon scattering to the electron mean free path, resulting in the reduction of energy acquired by electrons between the scattering events.

The linear slope of $\rho(J)$ also exhibits a significant dependence on the wire thickness, especially apparent at higher temperatures [Fig.~\ref{fig_3}(d)]. The slope increases with wire thickness up to 10 nm, and then decreases for Pt(15), in the temperature range up to $100$~K where the broadening was sufficiently small to allow a reliable determination of the slope. These nonmonotonic variations can be explained by the competition between the decrease of the phonon generation rate, with increasing Pt thickness, due to the smaller contribution of scattering at the Pt interfaces, and the increase of phonon escape time, as shown by the analysis below [see also inset in Fig.~\ref{fig_3}(d)].

To interpret the observed behaviors, and to evaluate the material parameters that control the non-equilibrium phonon distribution, we perform kinetic rate analysis of the current-driven phonon population. In the Drude-Sommerfeld approximation, the electron mean free path is $l_e=\frac{m^*v_F}{ne^2\rho}$, where $ v_F $ and $ m^* $ are the Fermi velocity and the effective mass, respectively~\cite{mott1958theory}. In the presence of electric field, the rate of electron scattering per unit volume is $r=\frac{J}{el_e}=\frac{ne\rho}{v_Fm^{*}}J $. Assuming that one phonon is generated in each scattering event, the rate of phonon generation per unit volume is $\frac{dn_{ph}}{dt}\vert_{gen}=r$. Relaxation due to the quasi-ballistic phonon escape from Pt~\cite{supplementary} can be described by the relaxation time approximation

\begin{equation}\label{rate_eq}
\frac{dn_{ph}}{dt}\vert_{rel}=-\frac{n_{ph}-n_{0}}{\tau_{ph}},
\end{equation}
where $ n_{0} $ is the phonon population in the absence of current, and $\tau_{ph}$ is the relaxation time, which is equal to the phonon escape time due to the rapid phonon escape. In the steady state,   $\frac{dn_{ph}}{dt}\vert_{gen}+\frac{dn_{ph}}{dt}\vert_{rel}=0$, or 
\begin{equation}\label{population}
n_{ph}=n_{0}+\frac{\tau_{ph} ne\rho}{v_{F}m^{*}}J.
\end{equation}
The parameters $n$, $v_{F}$, and $\rho$ in this expression are generally dependent on the current density $J$. For good metals such as Pt, the carrier density is to a very good approximation independent of current. To analyze the bias-driven variations of $v_F$, we note that electric bias results in a shift of the Fermi surface, affecting the average Fermi velocity~\cite{kittel1996introduction}. Using the Drude-Sommerfeld approximation, we find $\Delta \vec{v}=\vec{J}/ne$ for the bias-driven change of the electron velocity between the scattering events. For Pt(5) at $I=4$~mA, we calculate that $\Delta v_F$ is less than $5\%$ of $v_F$. Furthermore, the net effect is expected to become negligible when averaged over the Fermi surface, since the variation of the magnitude of electron velocity depends on the direction of wavevector. In contrast, the dependence of $\rho$
on $J$ is generally non-negligible, as is apparent from the experimental results discussed above. Below, we first derive the general expression accounting for this dependence, and then show that the contribution of the current-dependence of $\rho$ to the phonon population in our measurements is small. 

To establish the relationship between the current dependence of resistance and phonon generation/relaxation characteristics, we use the Matthiessen's rule for the electron mean free path in the presence of current-generated phonons, $1/l_{e}=1/l_{e,0}+n_{ph}\sigma_{e-ph}$. Here, $l_{e,0}$ is the mean free path in the absence of phonons, and $\sigma_{e-ph}$ is the average electron-phonon scattering cross section. Combining with Eq.~(\ref{population}), we obtain 
\begin{equation}\label{rhonl}
\rho(J)=\frac{\rho(0)}{1-\tau_{ph}\sigma_{e-ph} J/e}.
\end{equation}

Expanding in powers of the current density, we obtain to the lowest order in $\tau_{ph}\sigma_{e-ph}J/e$
\begin{equation}\label{rholin}
\rho(J)\approx\rho(0)[1+\tau_{ph}\sigma_{e-ph}J/e].
\end{equation}
According to this relation, to the lowest order in $J$, the resistivity is expected to depend linearly on the current density, in agreement with our experimental data. 
The validity of Eq.~(\ref{rholin}) is contingent upon several conditions. First, the kinetic rate equation Eq.~(\ref{rate_eq}) relies on negligible phonon thermalization. We utilized several independent approaches to estimate that this condition is well-satisfied in the studied wires~\cite{supplementary}. For very thick films or bulk samples, phonon thermalization is expected to result in the usual Joule heating characterized by a qualitatively different dependence $\rho(J)$. Second, the generated phonons must efficiently scatter electrons, which requires that their characteristic momentum is comparable to the Fermi momentum of electrons, such that the electron scattering on the generated phonons can described by the average scattering cross-section parameter $\sigma_{e-ph}$. Because of the energy conservation, this condition is not satisfied at small bias, consistent with the observed rounding of the linear dependence around $J=0$. Finally, the linear approximation for $\rho(J)$ holds only for $J\ll e/\tau_{ph}\sigma_{e-ph}$. Based on Eq.~(\ref{rholin}), this condition can be equivalently formulated as $(R(J)/R(0)-1)\ll 1$. This is well-satisfied for the the presented measurements, with the largest value of $(\rho(J)/\rho(0)-1)\approx 0.17 $ reached at $I=4$~mA for the Pt(5) wire. We have performed additional measurements of the nonlinear regime at larger currents, which could be well described by the general Eq.~(\ref{rhonl}) without any additional fitting parameters~\cite{supplementary}.

It should be emphasized that the obtained results are independent of the  functional form of $\rho(T)$, which usually exhibits an approximately linear dependence on temperature at high temperatures, and saturates at low  temperatures, due to the freeze-out of large-momentum phonons. As a consequence, Joule heating results in a quadratic or even slower dependence $\rho(J)$, instead of the linear dependence in the non-thermalized regime discussed above. 

These differences are closely related to the differences between phonon populations and characteristic energies in the two regimes. In particular, according to Eq.~(\ref{population}), the population of non-equilibrium phonons is proportional to current density. Meanwhile, their characteristic energy is determined by the energy acquired by electrons due to electric field between the scattering events, which is also approximately proportional $J$. We can contrast this with Joule heating at sufficiently high temperatures, when $\rho(T)$ is approximately linear, and $\rho(J)$ is quadratic. In the degenerate regime above the Debye temperature, the average phonon energy is independent of $T$, while according to the Rayleigh-Jeans law, their population is proportional to $T$. Thus, in the Joule heating regime, the average phonon energy is independent of $J$, while their population is quadratic in $J$. The total phonon energy  $\propto J^2$ is the same in both regimes, as expected since the dissipated power is $w=\rho J^2$.

The result Eq.~(\ref{rholin}) demonstrates that the linear slope of the dependence $\rho(J)$ provides direct information about phonon relaxation and electron-phonon scattering. For the studied Pt wires, because of the effects of electron scattering at the film interfaces, as reflected the thickness dependence of resistivity [inset in Fig.~\ref{fig_1}(a)], the phonon population [and thus the slope of $\rho(J)$], determined by the balance between phonon generation and escape rates, exhibits a complex dependence on $d$ [Fig.~\ref{fig_3}(d)]. This is reflected in Eq.~(\ref{rholin}) by the dependence of the slope of $\rho(J)$ on both $\rho(0)$ and $\tau_{ph}$. To gain insight into the observed variations, we use $\sigma_{e-ph}$ determined from the temperature dependence of resistivity~\cite{supplementary} and the measured $\rho(0)$ to calculate the values of $\tau_{ph}$ for different Pt wire thicknesses $d$. The dependence of $\tau_{ph}(d)$, determined from the $T=5$~K data, is well described by a linear function with zero intercept, as expected for the substrate-dominated relaxation  [inset in Fig.~\ref{fig_3}(d)]. The experimental values of $\tau_{ph}$ are in a semi-quantitative agreement with the calculation of the quasi-ballistic phonon escape time into the substrates based on the acoustic mismatch theory~\cite{supplementary}. These results confirm that phonon relaxation in the studied Pt wires is dominated by the fast quasi-ballistic phonon escape into the substrate, which facilitates non-equilibrium current-driven phonon distribution.

\begin{figure}
	\includegraphics[width=\columnwidth]{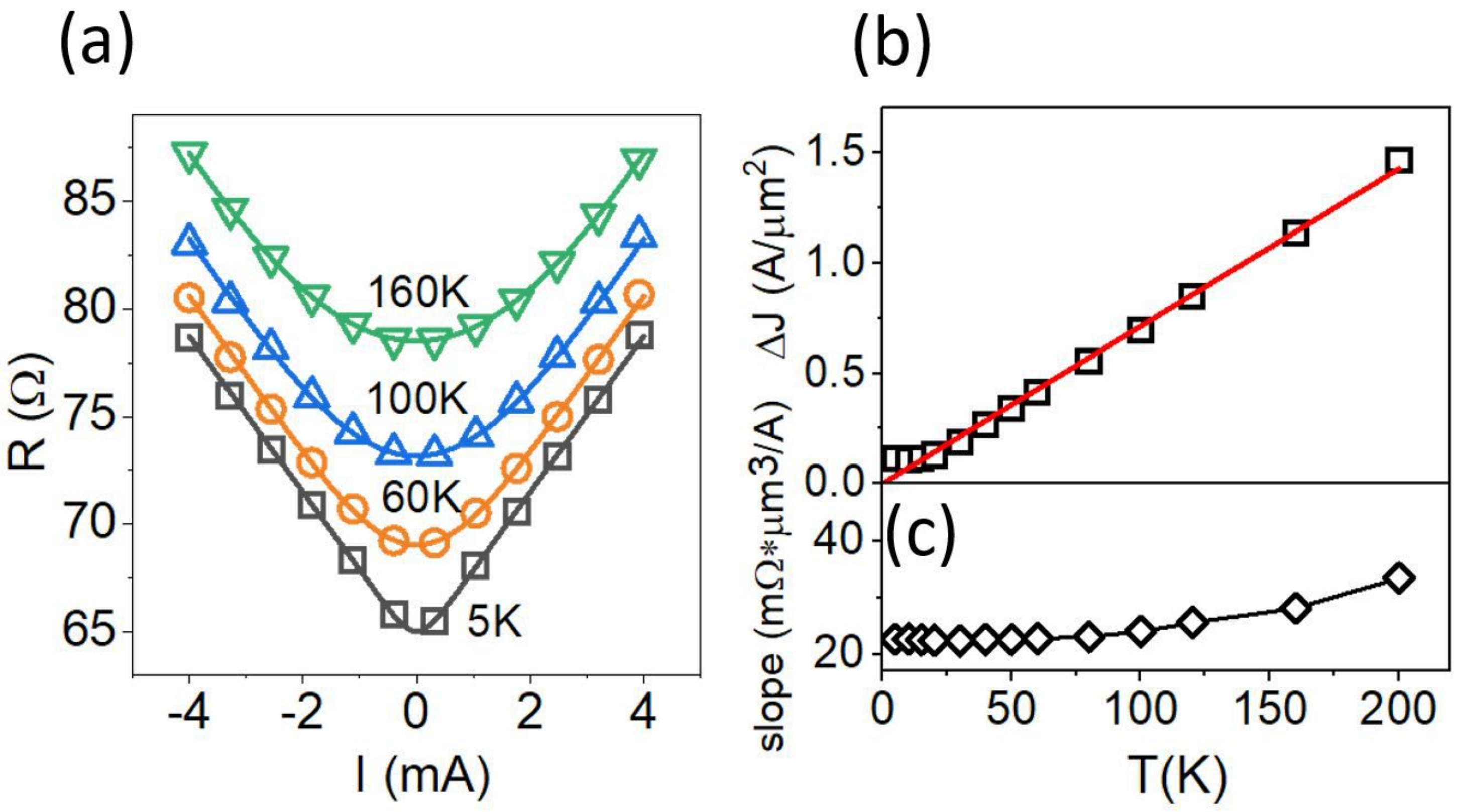}
	\caption{(Color online). (a) Symbols: resistance vs current for a $1$~$\mu$m-long, $500$~nm-wide Au(5) wire on Si substrate, at the labeled values of temperature. Curves: results of the data fitting with a linear function $R(I)=R(0)+\alpha|I|$ convolved with the Gaussian. (b),(c) Parameters extracted from the data fitting: the Gaussian width $\Delta I$ (b) and the slope of the linear dependence (c). The line in (b) is the best linear fit of the data for $T>20$~K.}\label{fig_4}
\end{figure}

We have confirmed the general relevance of the observed behaviors to nanostructures with efficient thermal dissipation, by measurements of current-dependent resistance in Pt wires fabricated on different substrates, as well as resistive metallic nanocontacts~\cite{supplementary}. Here, we briefly discuss the results for an Au(5) wire deposited on HF-cleaned undoped Si substrate. A Ni(0.5) wetting layer was inserted between Au(5) and the substrate, to improve adhesion and ensure the continuity of the ultrathin Au(5). The geometry of this wire was identical to that of the Pt(5) wire in Figs.~\ref{fig_1},~\ref{fig_2}. At $T=5$~K, the dependence $R(I)$ is well approximated by the linear function [Fig.~\ref{fig_4}(a)], in agreement with the results for Pt. This dependence becomes increasingly broadened with increasing temperature, with the broadening proportional to temperature at $T>20$~K  [ Fig.~\ref{fig_4}(b)]. The temperature dependence of the broadening is almost identical to that for Pt(5). At $T=5$~K, the resistance of the Au(5) wire is $4$ times smaller than that of Pt, while the linear slope is about $3.5$ times smaller than for Pt(5). In contrast to Pt(5), the slope for Au(5) monotonically increases with temperature. This dependence is similar to that observed for Pt(10) and Pt(15), but the magnitude of the variations is smaller, closer to Pt(5) and Pt(10).

The similarities between the results for Pt and Au, two materials with very different electronic band structures, confirms that the mechanisms underlying the observed effects are likely quite general to micro- and nano-scale systems characterized by efficient thermal dissipation. The almost identical thermal broadening for Pt and Au implies that the energy acquired by the electrons due to the electric field between the scattering events is almost the same for the two materials. Based on the Drude-Sommerfeld approximation, this energy is proportional to the effective electron mass and Fermi velocity, and inversely proportional to the electron density. The latter is similar in Pt and Au, while the effective electron mass is about $1.6$ times larger in Pt and the Fermi velocity is about $1.4$ times smaller~\cite{PhysRevB.2.4813}. Thus, the almost identical values of $\Delta J$ likely result from a fortuitous cancellation of different contributions to thermal broadening. Studies of other materials where such a fortuitous cancellation is not expected can provide a critical test for the proposed interpretation.

At $T=5$~K, the normalized slope $\frac{1}{\rho(0)}\frac{d\rho(J)}{dJ}$ is only $20\%$ smaller for Pt(5) than for Au(5). According to Eq.~\ref{rholin}, this implies that $\tau_{ph}\sigma_{e-ph}$ is similar for these materials. However, estimates based on the temperature dependence of resistivity show that  $\sigma_{e-ph, Pt}\approx 4\sigma_{e-ph, Au}$, and therefore the phonon relaxation time in Au(5) at $5$~K is about $5$ times larger than in Pt(5). The difference between phonon relaxation times becomes even larger at higher temperatures, as manifested by the increase of the slope for Au(5). This difference is likely associated with the much larger phonon escape time from the Au film. The escape time calculated from the acoustic mismatch at Au/Si interface, assuming isotropic momentum distribution of the generated phonons, is $75$~ps, less than half of the relaxation time determined from the slope of $R(I)$ at $5$~K~\cite{supplementary}. We speculate that the large value of $\tau_{ph}$ for Au originates from its poor wetting of the substrate, resulting in reduced phonon transparency of the interface with Si. Elucidating the relationship between wetting and phonon transparency of interfaces, by measurements such as those presented here, will be important for gaining further insight into thermal relaxation mechanisms in nanostructures, and for optimizing the thermal management in nanoelectronic devices.

In summary, our measurements and analysis have demonstrated that the phonons generated by electric current in conducting microstructures characterized by efficient thermal dissipation, generally form a strongly nonequilibrium distribution that cannot be described by a temperature. The nonequilibirum distribution is manifested by the linear dependence of resistance on current qualitatively inconsistent with the expected effects of Joule heating. The linear dependence is observed at sufficiently high currents at temperatures as high as $200$~K. The dependence of the observed phenomena on the structure, geometry, substrate and interface properties, and temperature can provide unique information about the electron-phonon scattering, phonon relaxation rates and mechanisms, and thermal effects. 

The most important message of our work is that on the microscopic level, scattering of electrons driven by electric bias generates phonons with a non-thermal distribution, whose contribution to resistance and other material properties can be qualitatively different from those expected for Joule heating. This result is expected to have broad implications for the optimization of thermal properties of electronic devices and nanostructures, and for the studies of current-induced phenomena. For instance, at high current densities, the linear dependence $R(I)$ associated with the non-equilibrium phonon distribution can fall significantly below the approximately quadratic dependence expected for Joule heating. In this regime, the average energy of phonons generated by current is larger than that of thermal phonons described by Joule heating, but their population is smaller than the thermalized population with the same total energy. This regime may be advantageous for the optimization of thermal management in nanoscale devices, for several reasons. First, additional energy dissipation due to electron scattering on the generated phonons can be significantly smaller than in the Joule heating limit, reducing the possibility of thermal runaway~\cite{PhysRevB.76.085433}. Second, the escape of the generated nonequilibrium phonons from the system can be more efficient than for Joule heating, due to the smaller rates of phonon-phonon scattering. Finally, the generated high-frequency phonons are less likely to contribute to current-induced physical degradation associated with the slow metastable mechanical degrees of freedom.

Current-induced phenomena have been extensively studied in the context of spin Hall effect, in thin Pt films similar to those discussed in our work~\cite{PhysRevLett.106.036601,jungwirth2012spin,PhysRevLett.101.036601,PhysRevLett.110.147601,demidov2014nanoconstriction,yang2016spin}. Our results may warrant re-examination of the Joule heating effects in these experiments. Similarly, low-temperature thermoelectric measurements at nanoscale commonly employ resistive heating of wires similar to those analyzed in our study. As we have shown, such resistive heating can result in a strongly nonequilibrium distribution characterized by substantially different phonon density (and its gradients) than inferred from Joule heating. We also note that the nonequilibrium phonon effects discussed above may have contributed to the linear current dependence of resistance recently observed at cryogenic temperatures in nanoscale spin valves, and attributed to quantum magnetization fluctuations~\cite{PhysRevLett.119.257201}.

This work was supported by the U.S. Department of Energy (DOE), Basic Energy Sciences (BES), under Award \# DE-SC2218976.

\bibliography{bibl0110}{}

\end{document}